\documentstyle[12pt]{article}
\begin{document}
\begin{center}
{\large\bf Research and Education in Basic Space Science}\\
{\bf The Approach Pursued in the UN/ESA Workshops}\\[0.5cm]

HAMID M.K. Al-NAIMIY$^a$ (Jordan), CYNTHIA P. CELEBRE$^b$ (Philippines), 
KHALIL CHAMCHAM$^c$ (Morocco), H.S. PADMASIRI DE ALWIS$^d$ (Sri Lanka), 
MARIA C. PINEDA DE CARIAS$^e$ (Honduras), HANS J. HAUBOLD$^f$ (United Nations), 
ALEXIS E. TROCHE BOGGINO$^g$ (Paraguay)
\end{center}
\noindent
ABSTRACT. Since 1990, the United Nations in cooperation with the European Space Agency  is holding annually a workshop on basic space science for the benefit of the worldwide development of astronomy. These workshops have been held in countries of Asia and the Pacific (India, Sri Lanka), Latin America and the Caribbean (Costa Rica, Colombia, Honduras), Africa (Nigeria), Western Asia (Egypt, Jordan), and Europe (Germany, France). Additional to the scientific benefits of the workshops and the strengthening of international cooperation, the workshops lead to the
establishment of astronomical telescope facilities in Colombia, Egypt, Honduras, Jordan, Morocco, Paraguay, Peru, Philippines, Sri Lanka, and Uruguay. The annual UN/ESA Workshops continue to pursue an agenda to network these astronomical telescope facilities through similar research
and education programmes. Teaching material and hands-on astrophysics material has been developed for the operation of such astronomical telescope facilities in an university environment.\par
\bigskip
\noindent
{\bf 1. Introduction}

          Research and education in astronomy and astrophysics are an international enterprise and the astronomical community has long shown leadership in creating international collaborations and cooperation: Because (i) astronomy has deep roots in virtually every human culture, (ii) it helps to understand humanity's place in the vast scale of the universe, and (iii) it teaches humanity about its origins and evolution.  Humanity's activity in the quest for the exploration of the universe is reflected in the history of scientific institutions, enterprises, and sensibilities. The institutions that sustain science; the moral, religious, cultural, and philosophical sensibilities of scientists themselves; and the goal of the scientific enterprise in different regions on Earth are subject of intense study (Pyenson and Sheets-Pyenson 1999).     

          The Bahcall report for the last decade of the 20th century (Bahcall, 1991) has been prepared primarily for the North American astronomical community, however, it may have gone unnoticed that this report had also an impact on a broader international scale, as the report can be used, to some extend, as a guide to introduce basic space science, including astronomy and astrophysics, in nations where this field of science is still in its infancy. Attention is drawn to the world-wide-web site at http://www.seas.columbia.edu/$\sim$ah297/\\un-esa/ where the initiative is publicized on how developing nations are making efforts to introduce basic space science into research and education curricula at the university level. This initiative was born in 1990 as a collaborative effort of developing nations, the United Nations (UN), the European Space Agency (ESA), and the Government of Japan, and covers the period of time of the last decade of the 20th century. Through annual workshops and subsequent follow-up projects, particularly the establishment of astronomical telescope facilities, this initiative is gradually bearing results in the regions of Asia and the Pacific, Latin America and the Caribbean, Africa, and Western Asia.\par 
\bigskip
\noindent
{\bf 2. Workshops on Basic Space Science}

          In 1959, the United Nations recognized a new potential for international cooperation and formed a permanent body by establishing the Committee on the Peaceful Uses of Outer Space (COPUOS). In 1970, COPUOS formalized the UN Programme on Space Applications to strengthen cooperation in space science and technology between non-industrialized and industrialized nations. In 1991 the UN, in close cooperation with developing nations, ESA, and the Government of Japan, started under the auspices of COPUOS a series of annual Workshops on Basic Space Science, which were hosted by UN member States in the five economic regions defined by the UN: India, Costa Rica, Colombia, Nigeria, Egypt, Sri Lanka, Germany, Honduras, Jordan, and France (Haubold 1998). 

          Over the past ten years, the workshops established  a close interaction between scientists from developing  and industrialized nations to discuss research findings at the current front lines in basic space science. The workshops also initiated a direct interaction between scientists from developing nations.  In depth discussions in working groups were fostered to allow the identification of the needs - especially common needs, which could be addressed on a larger scale - to enhance the participation of the developing nations in basic space science and to identify the best ways and means in which each nation could accelerate its participation in a meaningful endeavor. 

          The eighth in the series of UN/ESA Workshops on Basic Space Science, which, among other topics, addressed the feasibility of establishing a World Space Observatory (WSO), was held at Jordan in 1999. The ninth workshop will be held at France in 2000 and preparations are ongoing for the eleventh workshop to be held in Mauritius in 2001. The UN, ESA, Japan, and international organizations  will continue to provide assistance for the establishment and operation of astronomical facilities in Colombia, Egypt, Honduras, Jordan, Morocco, Paraguay, Peru, the Philippines, Sri Lanka, and Uruguay. \par
\bigskip
\noindent
{\bf 3. Astronomical Telescope Facilities}

          A number of Governments (among them Honduras 1997 and Jordan 1999), in cooperation with international partners, have acquired and established astronomical telescope facilities in their countries (Meade 16" Schmidt-Cassegrain models).  

          In conjunction to the workshops, to support research and education in astronomy, the Government of Japan has donated high-grade equipment to a number of developing nations (among them Sri Lanka 1995, Paraguay 1998, the Philippines 2001) within the scheme of ODA of the Government of Japan (Kitamura 1999). We refer here to 45cm high-grade astronomical telescopes furnished with photoelectric photometer, computer equipment, and spectrograph (or CCD). After the installation of the telescope facility by the host country and Japan, in order to operate such high-grade telescopes, young observatory staff members from Sri Lanka and Paraguay have been invited by the Bisei Astronomical Observatory for education and training, sponsored by the Japan International Cooperation Agency [JICA] (Kitamura 1999, Kogure 1999). 

          The research and education programmes at the newly established telescope facilities will focus on time-varying phenomena of celestial objects. The 45cm class reflecting telescope with photoelectric photometer attached is able to detect celestial objects up to the 12th magnitude and with a CCD attached up to the 15th magnitude, respectively. Such results have been demonstrated for the light variation of the eclipsing close binary star V505 Sgr, the X-ray binary Cyg X-1, the eclipsing part of the long-period binary e Aur, the asteroid No.45 Eugenia, and the eclipsing variable RT Cma Teles (Kitamura 1999). In forthcoming workshops, common observational programmes for variable stars for all the telescope facilities are envisaged.\par
\bigskip
\noindent
{\bf 4. Observing with the Telescopes: Research}

          In the course of preparing the establishment of the above astronomical telescope facilities, the workshops made intense efforts to identify available material to be used in research and education by utilizing such facilities. It was discovered that variable star observing by photoelectric or CCD photometry can be a prelude to even more advanced astronomical activity. Variable stars are those whose brightness, colour, or some other property varies with time. If measured sufficiently carefully, almost every star turns out to be variable. The variation may be due to geometry, such as the eclipse of one star by a companion star, or the rotation of a spotted star, or it may be due to physical processes such as pulsation, eruption, or explosion. Variable stars provide astronomers with essential information about the internal structure and evolution of the stars. The most predominant institution in this specific field of astronomy is the American Association of Variable Star Observers. AAVSO co-ordinates variable star observations made by amateur and professional astronomers, compiles, processes, and publishes them, and in turn, makes them available to researchers and educators.  The AAVSO receives over 350,000 measurements a year, from more than 550 observers world-wide. The measurements are entered into the AAVSO electronic database, which contains close to 10 million measurements of several thousand stars. 

          To facilitate the operation of variable star observing programmes and to prepare a common ground for such programmes, AAVSO developed a rather unique package titled ``Hands-On Astrophysics'' which includes 45 star charts, 31 35mm slides of five constellations, 14 prints of the Cygnus star field at seven different times, 600,000 measurements of several dozen stars, user-friendly computer programmes to analyze them, and to enter new observations into the database, an instructional video in three segments, and a very comprehensive manual for teachers and students (http://www.aavso.org/). Assuming that the telescope is properly operational, variable stars can be observed, measurements can be analyzed and send electronically to the AAVSO. 

          The flexibility of the ``Hands-On Astrophysics'' material allows an immediate link to the teaching of astronomy or astrophysics at the university level by using the astronomy, mathematics, and computer elements of this package. It can be used as a basis to involve both the professor and the student to do real science with real observational data. After a careful exploration of ``Hands-On Astrophysics'' and thanks to the generous cooperation of AAVSO, it was adopted by the above astronomical telescope facilities for their observing programmes (Mattei and Percy 1999, Percy 1991). The results of this effort will be reviewed at forthcoming workshops on basic space science.\par
\bigskip
\noindent
{\bf 5. Teaching Astrophysics: Education}

          Various strategies for introducing the spirit of scientific inquiry to universities, including those in developing nations, have been developed  and analyzed (Wentzel 1999a). What concerns the spirit of the workshops on basic space science, they have been organized and hosted by Governments and scientific communities which agreed beforehand on the need to introduce or further develop basic space science at the university level and to establish adequate facilities for pursuing such a field of science in practical terms, i.e., to operate an astronomical facility for the benefit of the university or research establishment (and to prospectively make the results from the facility available for public educational efforts). Additional to the hosting of the workshops, the Governments agreed to operate such a telescope facility in a sustained manner with the call on the international community for support and cooperation in devising respective research and educational programmes. Gradually, this policy is being implemented for those telescope facilities established through the workshops in cooperation with the UN, ESA, Japan, and other national and international organizations. 

          Organizers of the workshops have acknowledged in the past the desire of the local scientific communities to use educational material adopted and available at the local level (prepared in the local language). However, the workshops have also recommended to explore the possibility to develop educational material (additional to the above mentioned ``Hands-On Astrophysics'' package) which might be used by as many as possible university staff in different nations while preserving the specific cultural environment in which astronomy is being taught and the telescope is being used. A first promising step in this direction was made with the project ``Astrophysics for University Physics Courses'' (Wentzel 1999b). This project has been highlighted at the IAU/COSPAR/UN Special Workshop on Education in Astronomy and Basic Space Science, held during the UNISPACE III Conference at the United Nations Office Vienna in 1999 (Isobe 1999). Additionally, a number of text books and CD-ROMs have been reviewed over the years which, in the view of astronomers from developing nations, are particularly useful in the research and teaching process (for example, just to name three of them: Bennett et al. 1999, for teaching purposes; Lang 1999, a reference work in the research process; Hamilton 1996, a CD-ROM for astronomy in the classroom). This issue could be further discussed in the Newsletters, specifically published for the benefit of Africa (Querci and  Martinez 1999), Asia and the Pacific (Isobe 1999), and Latin America and the Caribbean (Eenens and Corral 1999).\par
\bigskip
\noindent 
{\bf 6. What Next?}

          To exactly take into account the obstacles encountered and the progress made, as observed in the ten-years long approach pursued in the UN/ESA Workshops as described above, the Ninth UN/ESA Workshop on Basic Space Science: Satellites and Telescopic Networks - Tools for Global Participation in the Studies of the Universe, hosted by the Centre National d'Etudes Spatiales (CNES) at the Observatoire Midi-Pyren\'{e}es (Universit\'{e} Paul Sabbatier), on behalf of the Government of France, 27-30 June 2000, Toulouse, France, will address the benefits of basic space science to society, particularly developing nations, and the experience with, results from, and the need for networks of astronomical telescopes in terms of common research and education programmes. The astronomical telescopes meant are particularly those mentioned above. During this workshop, additional working group sessions will be held to review the topics in sections 1 to 5 above and to chart the course for the future. 

          The workshop in France will also address in detail the feasibility to establish a World Space Observatory (WSO), discussed since the workshop in Sri Lanka in 1995, and the participation of developing nations in such an effort both from the point of view of research and education (Wamsteker and Gonzales Riestra 1997, United Nations GA Document A/AC.105/723).\par
\bigskip
\noindent
{\bf Acknowledgments}

          The authors are grateful to the following colleagues for sharing the pleasure to organize the UN/ESA Workshops on Basic Space Science and their follow-up projects as described in the above article: J. Andersen (IAU), J. Bennett (USA), S.C. Chakravarty (India), W. Fernandez (Costa Rica), S. Isobe (Japan), M. Kitamura (Japan), T. Kogure (Japan), K.R. Lang (USA), P. Martinez (South Africa), J. Mattei (AAVSO), P.N. Okeke (Nigeria), L.I. Onuora (UK), L. Pyenson (USA), F.-R. Querci (France), S. Rughooputh (Mauritius),  R. Schwartz (Germany),  M.A. Shaltout (Egypt), S. Torres\\ (Colombia), W. Wamsteker (ESA), and D.G. Wentzel (USA).\par
\bigskip
\noindent 
{\bf References}\par
\noindent
a) Astronomical Observatory, Higher Institute of Astronomy and Space Sciences, Al al-Bayt University, P.O. Box 130302, Al Mafraq, Jordan,\\ 
alnaimiy@yahoo.com\\
b) Astronomical Observatory, Philippine Atmospheric, Geophysical \& Astronomical Services Administration, 1424 Asia Trust Bank Building, Quezon Avenue, Quezon City, Philippines, cynthia$_-$celebre@hotmail.com\\
c) Faculty of Science Ain Chock, University Hassan II, B.P. 5366 Maarif, Casablanca 05, Morocco, chamcham@star.cpes.susx.ac.uk\\
d) Astronomical Observatory, Arthur C. Clarke Institut for Modern Technologies, Katubedda, Moratuwa, Sri Lanka, asela@slt.lk\\
e) Observatorio Astronomico, Universidad Nacional Autonoma de Honduras, Apartado Postal 4432, Tegucigalpa M.D.C., Honduras, mcarias@hondutel.hn\\
f) Office for Outer Space Affairs, United Nations, Vienna International Centre, P.O. Box 500, A-1400 Vienna, Austria, haubold@kph.tuwien.ac.at\\
g) Observatorio Astronomico, Facultad Politecnica, Universidad Nacional de Asuncion, Ciudad Universitaria, San Lorenzo, Paraguay, atroche@pol.com.py\par
\bigskip
\noindent
Bahcall, J.N., The Decade of Discovery in Astronomy and Astrophysics, National Academy Press, Washington D.C., 1991.\par
\smallskip
\noindent
Bennett, J., Donahue, M., Schneider, N., and Voit, M., The Cosmic Perspective, Addison Wesley Longman Inc., Menlo Park, California, 1999; a www site, offering a wealth of additional material for professors and students, specifically developed for teaching astronomy with this book and upgraded on a regular basis is also available: http://www.astrospot.com/.\par
\smallskip
\noindent
Eenens, Ph. and Corral, L., Astronomia Latino Americana (ALA), electronic Bulletin, http://www.astro.ugto.mx/$\sim$ala/.\par
\smallskip
\noindent 
Hamilton, C.J., Views of the Solar System CD-ROM, National Science Teachers Association, Arlington, 1996.\par
\smallskip
\noindent 
Haubold, H.J., ``UN/ESA Workshops on Basic Space Science: an initiative in the world-wide development of astronomy'',  Journal of Astronomical History and Heritage 1(2):105-121, 1998; an updated version of this paper is available at http://xxx.lanl.gov/abs/physics/9910042.\par
\smallskip
\noindent 
Isobe, S., Teaching of Astronomy in Asian-Pacific Region, Bulletin No. 15, 1999, published since 1991, isobesz@cc.nao.ac.jp.\par
\smallskip
\noindent 
Kitamura, M., ``Provision of astronomical instruments to developing countries by Japanese ODA with emphasis on research observations by donated 45cm reflectors in Asia'', in Conference on Space Sciences and Technology Applications for National Development: Proceedings, held at Colombo, Sri Lanka, 21-22 January 1999, Ministry of Science and Technology of Sri Lanka, pp. 147-152.\par
\smallskip
\noindent 
Kogure, T., ``Stellar activity and needs for multi-site observations'', in Conference on Space Sciences and Technology Applications for National Development: Proceedings, held at Colombo, Sri Lanka, 21-22 January 1999, Ministry of Science and Technology of Sri Lanka, pp. 124-131.\par
\smallskip
\noindent 
Lang, K.R., Astrophysical Formulae, Volume I: Radiation, Gas Processes and High Energy Astrophysics, Volume II: Space, Time, Matter and Cosmology, Springer-Verlag, Berlin 1999.\par
\smallskip
\noindent 
Mattei, J. and Percy, J. R. (Eds.), Hands-On Astrophysics, American Association of Variable Star Observers, Cambridge, MA 02138, 1998;\\ http://www.aavso.org/.\par
\smallskip
\noindent 
Percy, J.R. (Ed.), Astronomy Education: Current Developments, Future Cooperation: Proceedings of an ASP Symposium, Astronomical Society of the Pacific Conference Series Vol. 89, 1991.\par
\smallskip
\noindent 
Pyenson, L. and Sheets-Pyenson, S., Servants of Nature: A History of Scientific Institutions, Enterprises, and Sensibilities, W.W. Norton \& Company, New York, 1999.\par
\smallskip
\noindent 
Querci, F.-R. and P. Martinez (Eds.), African Skies/Cieux Africains, Newsletter, four issues published since 1997, http://www.saao.ac.za/$\sim$wgssa/.\par
\smallskip
\noindent 
United Nations Report on the UN/ESA Workshop on Basic Space Science, hosted by Al al-Bayt University, Mafraq, Jordan, on behalf of the Government of Jordan, A/AC.105/723, 18 May 1999.\par
\smallskip
\noindent 
Wamsteker, W. and Gonzales Riestra, R. (Eds.), Ultraviolet Astrophysics Beyond the IUE Final Archive: Proceedings of the Conference, held at Sevilla, Spain, 11-14 November 1997, European Space Agency SP-413, pp. 849-855.\par
\smallskip
\noindent 
Wentzel, D.G., ``National strategies for science development'', Teaching of Astronomy in Asian-Pacific Region, Bulletin No. 15, 1999a, pp. 4-10.\par
\smallskip
\noindent 
Wentzel, D.G., Astrofisica para Cursos Universitarios de Fisica, La Paz, Bolivia, 1999b, English language version available at\\ http://www.seas.columbia.edu/$\sim$ah297/un-esa/astrophysics.

\end{document}